# Celestial Mechanics and Polarization Optics of the Kordylewski Dust Cloud in the Earth-Moon Lagrange Point L5

# Part I. 3D Celestial Mechanical Modelling of Dust Cloud Formation

Judit Slíz-Balogh[1,2], András Barta[2,3] and Gábor Horváth[2,*]

1: Department of Astronomy, ELTE Eötvös Loránd University,
H-1117 Budapest, Pázmány sétány 1, Hungary

2: Environmental Optics Laboratory, Department of Biological Physics, ELTE Eötvös Loránd
University, H-1117 Budapest, Pázmány sétány 1, Hungary

3: Estrato Research and Development Ltd., H-1124 Némervölgyi út 91/c, Budapest, Hungary

*: Corresponding author, e-mail address: gh@arago.elte.hu

## Abstract

Since the discovery in 1772 of the triangular Lagrange points L4 and L5 in the gravitational field of
two bodies moving under the sole influence of mutual gravitational forces, astronomers found a
large number of minor celestial bodies around these points of the Sun-Jupiter, Sun-Earth, Sun-Mars
and Sun-Neptune systems. The L4 and L5 points of the Earth and Moon may be empty due to the
gravitational perturbation of the Sun. However, in 1961 Kordylewski found two bright patches near
the L5 point, which may refer to an accumulation of interplanetary particles. Since that time this
formation is called the Kordylewski dust cloud (KDC). Until now only a very few computer
simulations studied the formation and characteristics of the KDC. To fill this gap, we investigated a
three-dimensional four-body problem consisting of the Sun, Earth, Moon and one test particle,
1860000 times separately. We mapped the size and shape of the conglomeratum of particles not
escaped from the system sooner than an integration time of 3650 days around L5. Polarimetric
observations of a possible KDC around L5 are presented in the second part of this paper.

## 1. Introduction

In 1767 Euler discovered three unstable collinear points (L1, L2, L3) and in 1772 Lagrange found
two triangular points (L4, L5) in the gravitational field of two bodies moving under the sole
influence of mutual gravitational forces (Szebehely 1967). In the three-body problem of celestial
mechanics the L4 and L5 Lagrange points are stable in linear approximation, if the mass ratio $Q = m_{smaller}/m_{larger}$ of the two primaries is smaller than $Q^* = 0.0385$ (Murray & Dermott 1999).
Astronomers found a large number of minor celestial bodies around these points of the planets of
our Solar System and the Sun. The most well-known are the Greek and Trojan minor planets around
the L4 and L5 points of the Sun-Jupiter system (Schwarz *et al.* 2015, Schwarz & Dvorak 2012).
Minor planets have also been found around the triangular Lagrange points of the Sun-Earth (John *et
al.* 2015), Sun-Mars (Christou 2017) and Sun-Neptune systems (Sheppard &Trujillo 2006).
    What about the vicinities of the Lagrange points L4 and L5 of the Earth and Moon? Since
the mass ratio $Q = m_{Moon}/m_{Earth} = 0.0123$ of the Moon and Earth is smaller than $Q^* = 0.0385$, the L4

and L5 points are theoretically stable. Thus, interplanetary particles with appropriate velocities could be trapped by them. In spite of this fact, they may be empty due to the gravitational perturbation of the Sun. Taking into account the perturbation of the Sun, the orbits in the vicinity of the L5 point have been computationally investigated in two dimensions (Slíz *et al.* 2015, 2017). According to the results of these simulations, if test particles start from the vicinity of the L5 point, their motion will be chaotic. This chaos is transient, and there are many trajectories which do not leave the system even for $10^6$ days, and long-existing (for 30-50 years) islands form around L5. Thus, although the gravitational perturbation of the Sun really sweeps out many trajectories from the L5 point on an astronomical time scale, on a shorter time scale there are many long-existing trajectories too.

In 1961 Kordylewski found two bright patches near the L5 point, which may refer to an accumulation of dust particles (Kordylewski 1961). Since that time this hypothetic formation is called the Kordylewski dust cloud (KDC). Until now only a very few computer simulations studied the formation and characteristics of the KDC (Slíz *et al.* 2015, 2017, Salnikova *et al.* 2018). To fill this gap, we investigate here a three-dimensional four-body problem consisting of three massive bodies, the Sun, the Earth and the Moon (primaries) and a low-mass test (dust) particle, 1860000 times separately. Our aim was to map the size and shape of the conglomeratum of particles not escaped from the system sooner than a given integration time around L5. Polarimetric observations of a possible KDC around L5 are presented in the second part (Slíz-Balogh *et al.* 2018) of this paper.

## 2. Models and Methods

We used a three-dimensional barycentric four-body model consisting of the Sun, Earth, Moon and a test particle (called simply 'particle' further on) near the L5 point of the Earth-Moon system (Figure 1). The initial positions and velocities of the Sun, Earth and Moon were taken from the freely available NASA JPL database (https://ssd.jpl.nasa.gov/horizons.cgi) in ecliptic coordinate system with Cartesian coordinates relative to the Solar System Barycenter. These coordinates (and also the input coordinates of the L5 point) were converted to Sun-Earth-Moon-particle barycentric ecliptic ones. All calculations were performed in this 3D Sun-Earth-Moon-particle barycentric ecliptic coordinate system, while the representations were made in a geocentric ecliptic coordinate system for the sake of a better visualization.

Regarding that it is a 3D model, the computation of the coordinates of the L5 point happened as follows:

1. Based on the Moon's orbital data obtained from the NASA JPL database, the longitude of the ascending node and the inclination with respect to Earth equator were calculated.

2. The ecliptic coordinates of the Moon were converted into equatorial ones, then they were rotated twice: first with the longitude of the ascending node about the $z$ axis, then with the inclination relative to Earth equator into the equatorial plane about the $x$ axis.

3. In the equatorial plane, the position and velocity of the L5 point were calculated with rotating the Moon's coordinates by 60° clockwise.

4. Finally, these coordinates were rotated back (first with the inclination relative to the Earth equator into the equatorial plane about the $x$ axis, and then with the longitude of the ascending node about the $z$ axis).

The potential energy $U$ and motion equations of the Sun-Earth-Moon-particle system are the following:

$$U = -\gamma \sum_{i=1, i \neq j}^{4} \frac{m_i m_j}{r_{ij}}, \qquad r_{ij} = \sqrt{\left(x_i - x_j\right)^2 + \left(y_i - y_j\right)^2 + \left(z_i - z_j\right)^2}, \qquad (1)$$

$$m_i \ddot{x}_i = -\frac{\partial U}{\partial x_i}, m_i \ddot{y}_i = -\frac{\partial U}{\partial y_i}, m_i \ddot{z}_i = -\frac{\partial U}{\partial z_i}, i = 1, 2, 3, 4, \qquad (2)$$

where $\gamma = 6.674 \cdot 10^{-11}$ m³kg⁻¹s⁻² is the universal gravitational constant, $m_1$, $m_2$, $m_3$ and $m_4$ are the masses of the Sun, the Earth, the Moon and the particle, $r_{12}$, $r_{13}$, $r_{14}$, $r_{23}$, $r_{24}$ and $r_{34}$ are the distances between the Sun-Earth, Sun-Moon, Sun-particle, Earth-Moon, Earth-particle and Moon-particle, respectively, in the *x-y-z* barycentric ecliptic coordinate system (Figure 1). The second-order non-linear differential equations (2) obtained by derivation from (1) were converted into a system of first-order differential equations, which were solved with an appropriate Runge-Kutta method. The motion equations were implemented in a dimensionless form where the characteristic length unit is 1 AU = 149597870.66 km and the time unit is 86 400 s (= 1 day). The computational method was an adaptive step size Runge-Kutta-Fehlberg 7(8) integrator (Fehlberg 1968) in which the actual step size is determined according to the desired accuracy $\varepsilon = 10^{-16}$ (= tolerated local error per unit step).

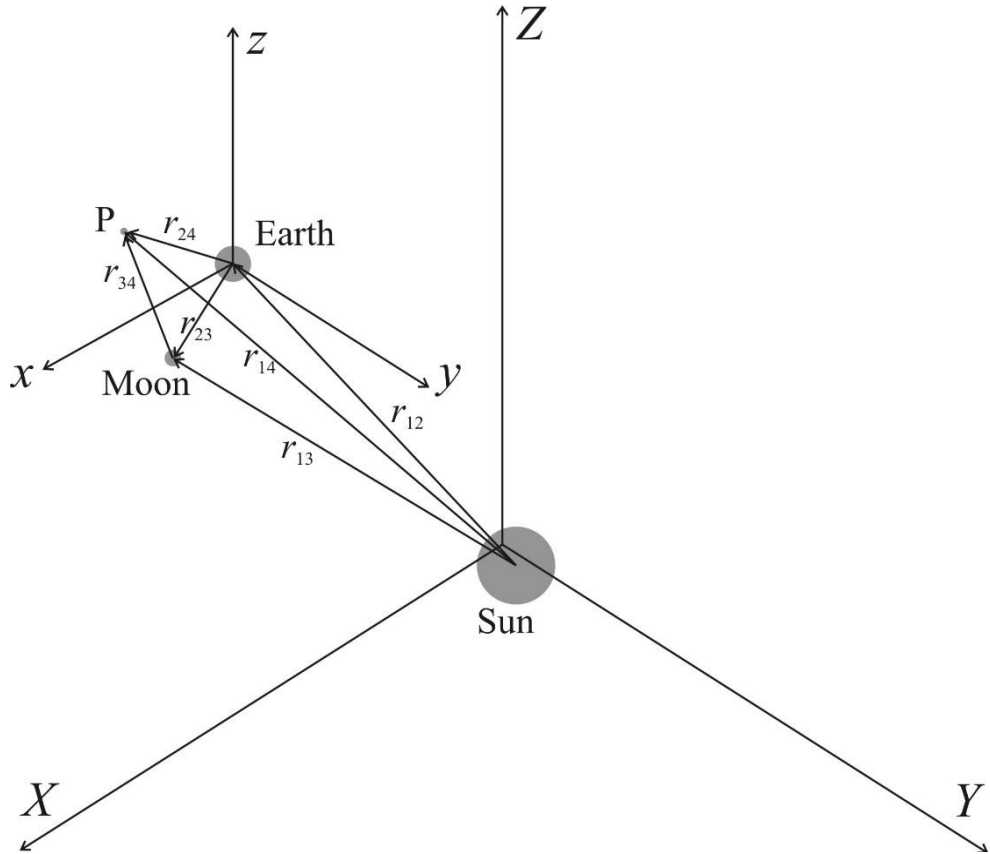

**Figure 1**: Position vectors of components (Sun, Earth, Moon, P: particle) of the studied four-body problem in the barycentric ecliptic coordinate system. The *x-y* plane is the plane of the ecliptic, the *x* axis points towards the vernal equinox, $\boldsymbol{r}_{12}$, $\boldsymbol{r}_{13}$, $\boldsymbol{r}_{14}$, $\boldsymbol{r}_{23}$, $\boldsymbol{r}_{24}$ and $\boldsymbol{r}_{34}$ denote the vectors of the Sun-Earth, Sun-Moon, Sun-particle, Earth-Moon, Earth-particle and Moon-particle, respectively. The particle is in the vicinity domain *V* around the L5 Lagrange point. The sizes and distances are not to scale.

A particle was considered as trapped (in fact, non-escaped), if in 3650 days it does not leave the spherical shell $0.5r_0 \leq D \leq 1.5r_0$ with $r_0 = \sqrt{(x_0 - x_{E0})^2 + (y_0 - y_{E0})^2 + (z_0 - z_{E0})^2}$, where $x_0$, $y_0$, $z_0$, $x_{E0}$, $y_{E0}$, $z_{E0}$ are the initial coordinates of the particle and the Earth at starting time $t_0$, and $D$ is the thickness of the shell. If we would like to model the KDC around the Lagrange point L5, then, of course, it is not enough starting the particles at a single $t_0$. In principle, continuous trapping should be modelled in a medium with unknown particle density and velocity. Instead of this, we used the following approach: For a given $t_0$, the motion equations were solved for 1860000 particles, the starting positions and velocities of which were distributed in the phase space in the vicinity domain $V$ of the L5 point (Table 1). The $V$ domain was divided into 41 equal parts in the $x$, $y$ and $z$ ranges, and 3 equal parts in the $v_x$, $v_y$ and $v_z$ ranges. The trapped (non-escaped) particles constitute a "particular dust cloud". This procedure was repeated 28 times for 28 different $t_0$-values. Finally, the obtained separate particular dust clouds were summed up resulting in the "summed dust cloud".

**Table 1**: The vicinity domain $V$. $x_{0,L5}$, $y_{0,L5}$, $z_{0,L5}$, $v_{x0,L5}$, $v_{y0,L5}$, $v_{z0,L5}$ denote the calculated initial position and velocity coordinates of the L5 Lagrange point at $t_0$. The size of the domain where the test particles are started is $0.0008 \times 0.0008 \times 0.0008$ in position and $0.00006 \times 0.00006 \times 0.000006$ in velocity range around L5. For units see text.

| $x_0$ | $y_0$ | $z_0$ |
|---|---|---|
| $\{x_{0,L5} - 0.0004,$ $x_{0,L5} + 0.0004\}$ | $\{y_{0,L5} - 0.0004,$ $y_{0,L5} + 0.0004\}$ | $\{z_{0,L5} - 0.0004,$ $z_{0,L5} + 0.0004\}$ |
| $v_{x0}$ | $v_{y0}$ | $v_{z0}$ |
| $\{v_{x0,L5} - 0.00003,$ $v_{x0,L5} + 0.00003\}$ | $\{v_{y0,L5} - 0.00003,$ $v_{y0,L5} + 0.00003\}$ | $\{v_{z0,L5} - 0.000003,$ $v_{z0,L5} + 0.000003\}$ |

In the motion equation (2) of the particle we considered only the gravitational forces of the three massive bodies and neglected forces induced by the radial solar radiation pressure and the Poynting-Robertson (P-R) drag. Here we show that this neglect was right. In an inertial reference system for a dust particle with geometric cross section $A$, the Poynting-Robertson drag force $\underline{F}_{P\text{-}R}$ and the radiation pressure force $\underline{F}_{rad}$ can be written as follows (Burns *et al.* 1979, Liou *et al.*1995):

$$\underline{F}_{P\text{-}R} = -\frac{SA}{c^2}Q_{pr}\underline{v}, \tag{3}$$

$$\underline{F}_{rad} = \frac{SA}{c}Q_{pr}\left(1 - \frac{\dot{r}}{c}\right)\underline{\dot{s}}, \tag{4}$$

where $\underline{v}$ is the velocity vector of the particle, $S$, $c$ and $Q_{pr}$ are the solar energy flux density, the speed of light in vacuum and the radiation pressure coefficient, respectively, while $\underline{\dot{s}}$ is the unit vector in the direction of the Sun, and $\dot{r}$ is the particle's radial velocity. Burns *et al.* (1979) and Liou *et al.* (1995) investigated the ratio of the radiation pressure force and the solar gravitation force and stated that for particles with sizes greater than a few μm the radiation pressure force is negligible. On the other hand, the very small particles (with radius less than 0.01 μm) are unaffected by the solar radiation, because the characteristic radiation wavelength is relatively too large to sustain absorption or scattering. The radiation pressure is maximal (and for certain materials may exceed the gravitational force) for particles with radius from 0.1 to 0.5 μm. Therefore, we calculated the gravitational force $F_{gr}$ of the Sun, Earth and Moon and the sum of the Poynting-Robertson drag force $\underline{F}_{P\text{-}R}$ and the radiation pressure force $\underline{F}_{rad}$ around this particle size range.

## 3. Results

In the case of a particle with medium density $\rho$ = 3000 kg/m$^3$ the value of $Q_{pr}$ is 1 (Burns *et al.* 1979). The solar energy flux density at one AU from the Sun is $S$ = 1.361 kW/m². Using these numerical values, we started our simulation at 01:14 (UT) on 19 August 2017 (Figure 2) and calculated the mentioned forces. According to Figure 2, over the particle mass $m$ = 10$^{-14}$ kg and particle radius $r$ = 1 μm, the gravitational force dominates, which is consistent with the earlier finding of Burns *et al.* (1979).

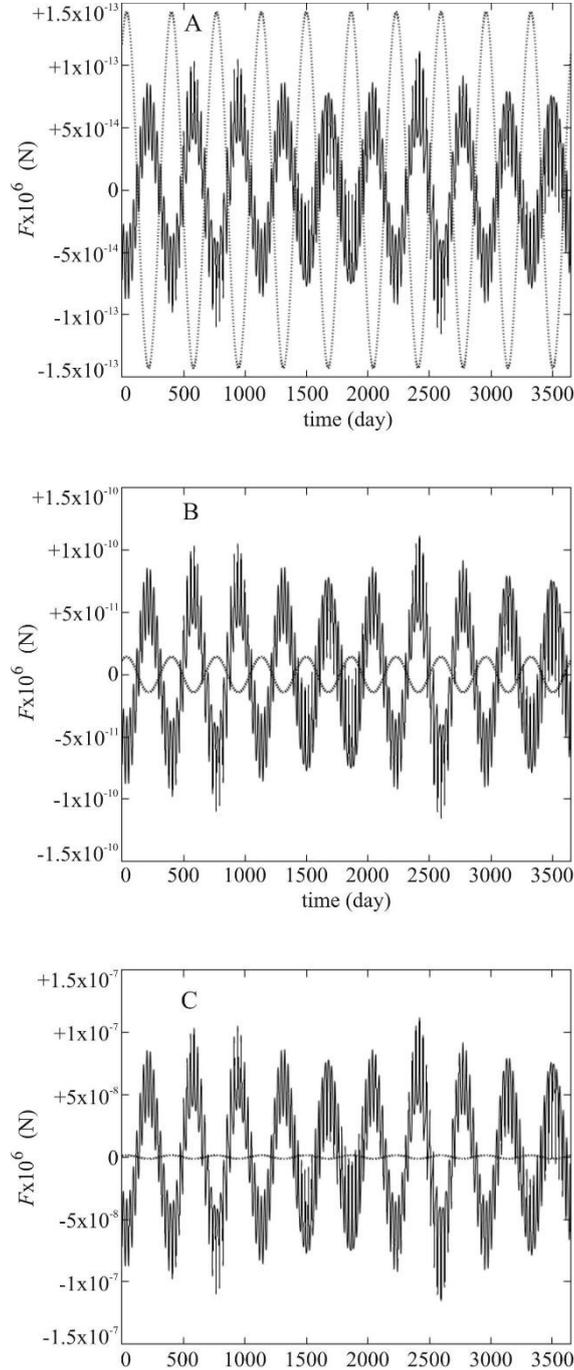

**Figure 2**: The gravitational force $F_{gr}$ of the Sun, Earth and Moon (continuous curve) and the sum of the Poynting-Robertson drag force $\underline{F}_{P\text{-}R}$ and the radiation pressure force $\underline{F}_{rad}$ (dotted curve) in $x$ direction for a particle with mass $m$ and radius $r$. (A) $m$ = 10$^{-17}$ kg, $r$ = 0.1 μm, (B) $m$ = 10$^{-14}$ kg, $r$ = 1 μm, (C) $m$ = 10$^{-11}$ kg, $r$ = 10 μm. The situation is similar in the $y$ direction, while in the $z$ direction $F_{gr}$ always is larger than $\underline{F}_{P\text{-}R}$ + $\underline{F}_{rad}$.

Hence, we obtained that both the radiation pressure and the P-R drag are negligible relative to the gravitational force for particles with radius over about 1 μm, while particles with radius between 0.1 and 0.5 μm can be ejected from the vicinity area of the L5 point after a long enough period (Burns *et al.* 1979). Thus, it was right not to take into consideration the radiation pressure and the P-R drag forces in our short-term simulations.

For a better understanding of the properties of the trapped particles around the L5 point, two types of simulations were performed. On the one hand, we studied the behaviour of a particular dust cloud versus time (Figure 3), and on the other hand, we investigated the summed dust cloud cumulated from 28 particular dust clouds (Figure 4).

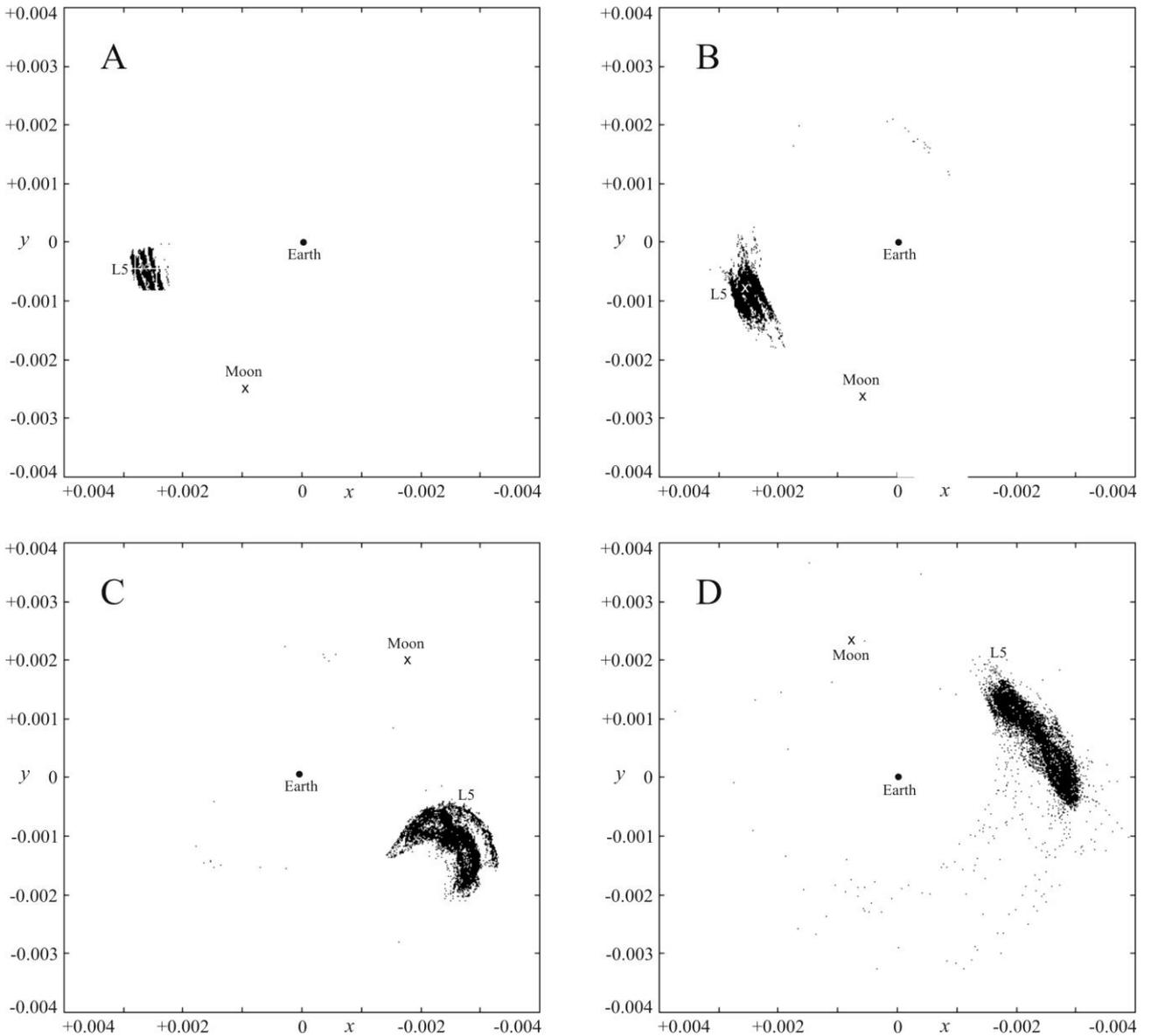

**Figure 3**: (A) Initial positions (black pixels) of the non-escaped trajectories of 1860000 particles started at $t_0 = 01:14$ (UT) on 22 August 2007 from the vicinity domain $V$ around the L5 point in geocentric ecliptic coordinate system. (B-D) The positions (black pixels) of these particles (composing a particular dust cloud) after 28 (B), 1460 (C) and 3650 days (D). Earth: dot (center of the picture), L5 point: ×, Moon: ×. A given black pixel means that in that direction of view there is at least one particle.

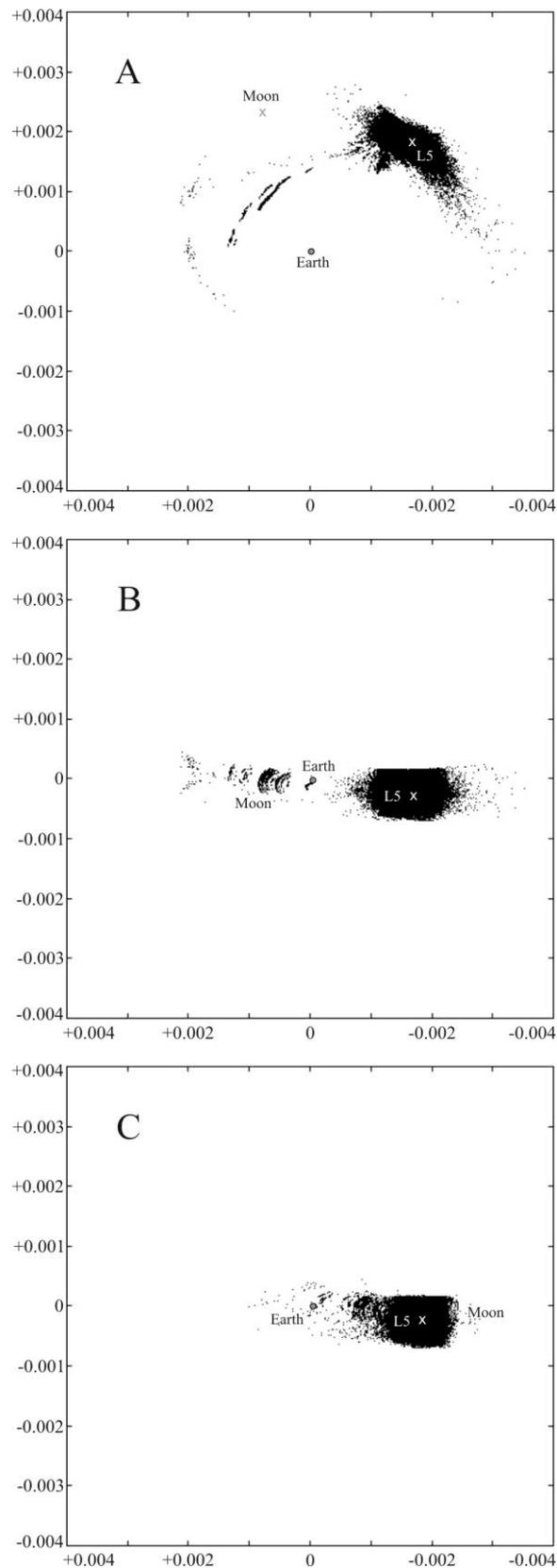

**Figure 4**: The summed dust cloud at 01:14 (UT) on 19 August 2017 in geocentric ecliptic coordinate system in the *x-y* (A), *x-z* (B) and *y-z* (C) plane. The *x* axis points towards the vernal equinox. The summed dust cloud is the combination of 28 separate particular dust clouds. The simulation of the 1st and 28th (last) particular dust cloud started at $t_0 = 01:14$ on 22 July and 01:14 on 19 August 2017, respectively. Earth: grey dot (center of the picture), L5 point: white ×, Moon: grey ×. A given black pixel means that in that direction of view there is at least one particle.

Figure 3A shows the initial positions of the non-escaped particles (particular dust cloud) for 3650 days around L5 started at $t_0 = 01:14$ (UT) on 22 August 2007 (Table 2) from the vicinity domain *V*. The particles started with different velocities compose a band structure.

**Table 2**: Initial positions and velocities of the the Sun, Earth, Moon and the Lagrange point L5 for the epoch JD = 2454334.551388889 ($t_0 = 01:14$ on 22 August 2007) in the solar system barycentric ecliptic coordinate system.

|  | **Sun** | **Earth** | **Moon** | **L5** |
|---|---|---|---|---|
| $x_0$ | 9.890645064385753 $\cdot 10^{-4}$ | 8.634609584780867 $\cdot 10^{-1}$ | 8.625611534079515 $\cdot 10^{-1}$ | 8.600146967391559 $\cdot 10^{-1}$ |
| $y_0$ | 4.795511121657655 $\cdot 10^{-3}$ | -5.238211225876513 $\cdot 10^{-1}$ | -5.263223250956695 $\cdot 10^{-1}$ | -5.255959724119573 $\cdot 10^{-1}$ |
| $z_0$ | -8.461861578803593 $\cdot 10^{-5}$ | -7.427675789134437 $\cdot 10^{-5}$ | -3.194245021219622 $\cdot 10^{-4}$ | -20.320528297276486 $\cdot 10^{-5}$ |
| $v_{x0}$ | -6.293333135136141 $\cdot 10^{-6}$ | 8.697706450830024 $\cdot 10^{-3}$ | 9.237525785409454 $\cdot 10^{-3}$ | 8.833349225711902 $\cdot 10^{-3}$ |
| $v_{y0}$ | 1.782035107187389 $\cdot 10^{-6}$ | 1.460718375982933 $\cdot 10^{-2}$ | 1.443790508519941 $\cdot 10^{-2}$ | 1.404098840128974 $\cdot 10^{-2}$ |
| $v_{z0}$ | 9.935326112885818 $\cdot 10^{-8}$ | -2.250105040215131 $\cdot 10^{-7}$ | 4.852769975236783 $\cdot 10^{-6}$ | -428.023679437153237 $\cdot 10^{-7}$ |

Figure 3B displays the particular dust cloud around L5 containing the same trapped particles, 28 days laterat 01:14 on 19 September 2007. The particles did not leave the vicinity of L5, the band structure is still discernible, but the particular dust cloud dispersed and became more homogeneous.

Figure 3C illustrates the particular dust cloud around L5 containing the same trapped particles, 1460 days (4 years) laterat 01:14 on 22 August 2011. The particles form an U-shaped diffuse cloud with two wings directed opposite to the direction of rotation. The L5 point is still within the cloud.

Figure3D depicts the particular dust cloud around L5 containing the same trapped particles, 3650 days (10 years) later at 01:14 on 19 August 2017. The L5 point is just at the leading edge of the elongated cloud having a long comet-like trail of broken-off particles. The ending (target) time 01:14 on 19 August 2017 of the simulations shown in Figures 3D and 4 was chosen to match the date of the imaging polarimetric patterns presented in the 2nd part (Slíz-Balogh, Barta, Horváth 2018) of this paper.

Figure 4 shows a summed dust cloud composed of 28 particular dust clouds formed daily between $t_0 = 01:14$ on 22 July and $t_0 = 01:14$ on 19 August 2017. Similarly to Figure3D, a comet-like trail of broken-off particles can be well observed. The particles forming some faint arcs were trapped from the edges of the vicinity domain *V*. The dimension of the summed dust cloud in the *z* direction is approximately half of that in the *x* or *y* direction.

Figure 5 shows the grey-coded particle density of the summed dust cloud. Depending on how many days earlier the particular dust cloud was trapped before the target date, two different types of structure can be distinguished: (i) Figure 5A shows the summed dust cloud at the target date 01:14 on 19 August 2017 with particles trapped 1-28 days earlier. (ii) Figure 5B shows the summed dust cloud at the target date 01:14 on 19 August 2017 with particles trapped 1-5, 8-18 and 21-28 days earlier. This cloud has a characteristic band structure around the L5 point with at least 6 bands. (iii) Figure 5C displays the summed dust cloud with the particles of which were trapped 6, 7, 19 and 20 days earlier. The cigar-like/elongated shape of this cloud is totally different from that in Figure 5A.

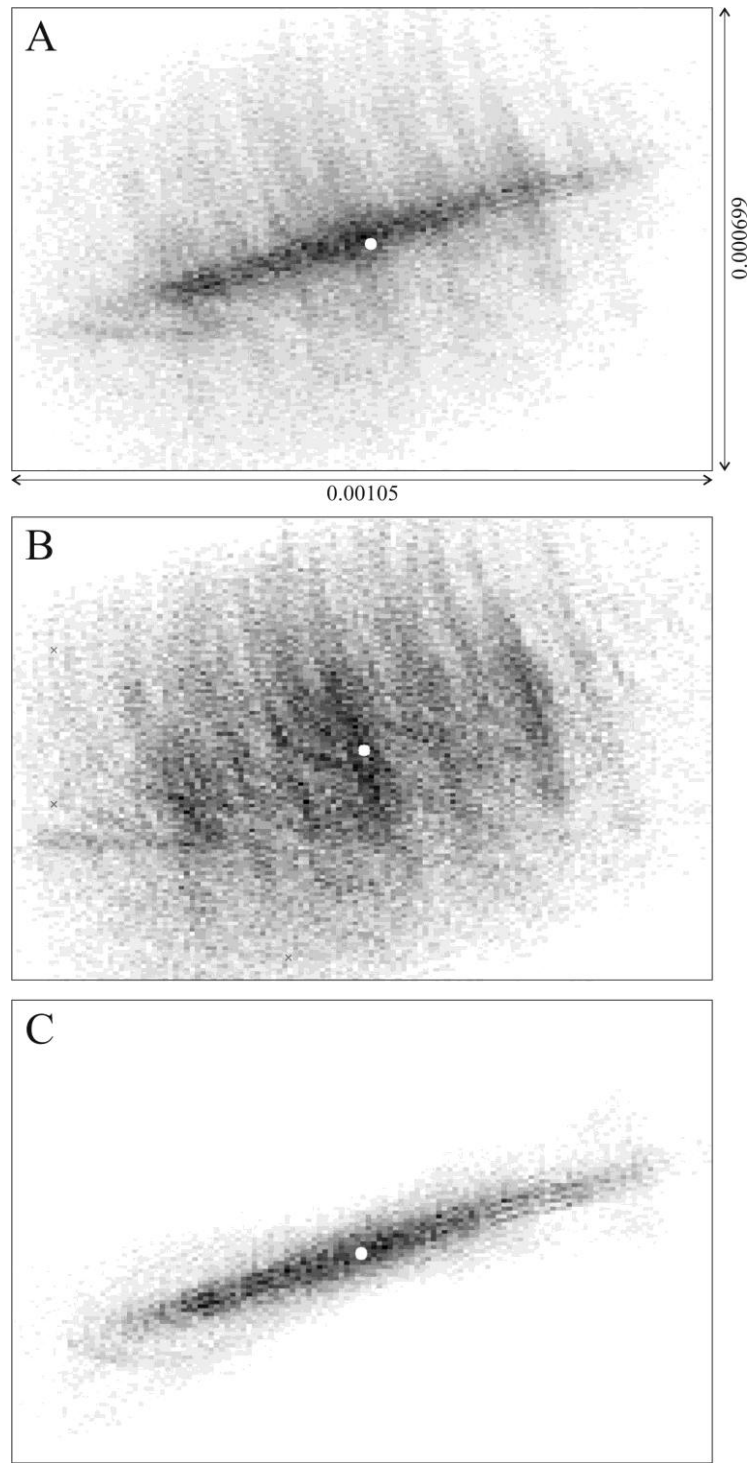

**Figure 5**: Computer-simulated density distribution of the particles of the KDC around the L5 point (white dot) of the Earth-Moon system in equatorial coordinate system as we would see in the sky. The angular extension of the picture is 22.5° (horizontal) × 15° (vertical). The horizontal and vertical axis denotes the direction of the right ascension (RA) and the declination (DE), respectively. (A) Summed dust cloud (at target date 01:14 on 19 August 2017) of the particular dust clouds, the particles of which were trapped 1-28 days earlier. (B) As (A) for particular dust clouds, the particles of which were trapped 1-5, 8-18 and 21-28 days earlier. (C) As (A) for particular dust clouds, the particles of which were trapped 6, 7, 19 and 20 days earlier. The darker the grey shade, the larger is the particle density.

The phenomenon of contraction is better seen in Figure 6 showing the contraction of three particular dust clouds (consisting of 927 particle's orbits) in the $z$ direction about 6-7 and 19-20 days after the starting time (23:00 on 28 July, 1 and 7 August 2017). The orbit of each particle is sinusoidal also in the $z$ direction with the same period as that in directions $x$ and $y$ coinciding with the lunar orbital period. In the case of all three particular dust clouds the same phenomenon has been experienced.

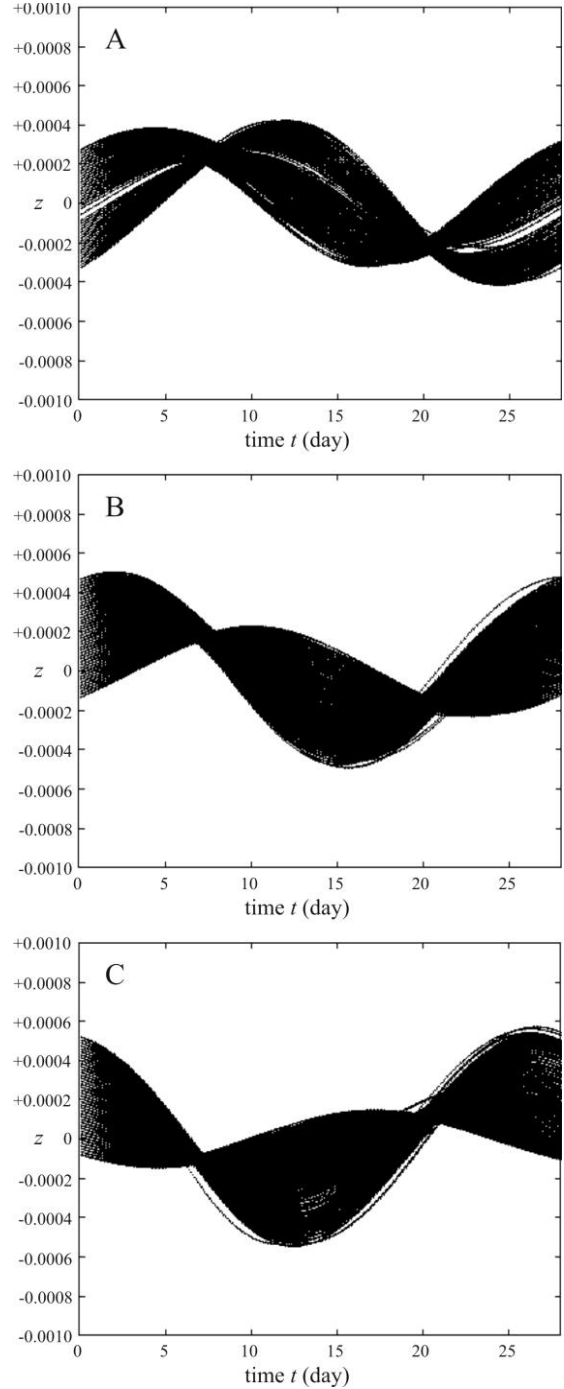

**Figure 6**: The $z$ coordinates of particles of three particular dust clouds with different starting times $t_0$ versus time $t$ from $t = 0$ to 28 days in geocentric ecliptic coordinate system (Figure 1). (A) $t_0 = 23:00$ (UT) on 28 July 2017, (B) $t_0 = 23:00$ on 1 August 2017, (C) $t_0 = 23:00$ on 7 August 2017. In all three cases, in the $z$ direction two contractions can be clearly seen after about 6-7 and 19-20 days. The particular dust cloud in B on 8-9 and 20-21 August shows an elongated, dense shape (Figure 4C), on the other days it has a less dense, banded form (Figure 5B).

## 4. Discussion

In special cases there are exact well-known classic analytical solutions of the three-body problem (Szebehely 1967, Rajnai *et al.* 2014). Recently, a new exact solution of a special case of the four-body problem was discovered (Érdi & Czirják 2017). However, the general and especially the three-dimensional four-body problem can be solved only numerically.

The stability of the L4 and L5 Lagrange points of the Earth and Moon has some well-exploitable advantages. They are suitable for spacecraft, satellite or space telescope parking with minimal fuel consumption (nonetheless at the moment there are no spacecraft orbiting neither at L4 nor at L5 in the Solar System), or they can be applied as transfer stations for the mission to Mars or other planets, and/or to the interplanetary superhighway. The investigation of the dynamics of the Earth-Moon Lagrange points is important as well from the point of view of space navigation safety. Since in the study of these points the gravitational effect of the Sun cannot be ignored, one has to study computationally a four-body problem, as we have done.

Figures 3 and 4 depict well the size and shape of the particular and summed dust clouds, but they do not give any information about the particle density. Therefore we created images of the summed dust cloud (Figure 5) where the picture area is uniformly divided into cells in the line of sight, and these cells are shaded with different grey hues depending on the number of particles in the cells. The structure of the summed dust cloud (Figure 5A) consists of two distinct parts: (i) an extended, less dense banded conglomeratum (Figure 5B), and (ii) an elongated denser one (Figure 5C). The length of the bands of a particular dust cloud varies periodically (synchronous with the Moon's orbital period) depending on how many days earlier the particles were trapped. After being trapped, the particular dust cloud begins to contract in the band direction, and about 6-7 days later its length is minimal and its densitiy is maximal. Then it starts to expand again, and reaches its maximal length after about other 6-7 days. If the trapping happens 6, 7, 19 or 20 days earlier, the elongated and dense particular clouds will dominate (Figures 5A and 5C). If about 6, 7, 19 or 20 days earlier there was not trapping, the summed dust cloud will look like shown in Figure 5B.

The lunar orbit's inclination with respect to the Earth equator on 19 August 2017 is 19.4°, which is the same as the angle of the elongated cloud's axis from the horizontal (Figure 5C). The angle of the band's axis from the vertical (Figure 5B) is the same as the angle of the equator relative to the ecliptic (23.44°). This means that the bands of conglomeratum are perpendicular to the ecliptic, while the elongated cloud is parallel to the lunar orbit plane, that is while the cloud is contracting, it is twisting/slewing, too.

The phenomenon described in Figure 6 is the periodic contraction of a particular dust cloud in the $z$ direction in ecliptic coordinate system. Given that the motion of each particle of the dust cloud is chaotic (Slíz *et al.* 2015, 2017), supposedly a chaotic set is in the background of this phenomenon, but its explanation is not the subject of this paper.

Our simulations assumed steadily discontinuous material capture. But in reality it is far from being so. For example, in the case of a meteor shower the amount of trapped particles is larger, while at other times it may be much smaller. So the bands in Figure 5, which are the results of trappings of different days earlier with different velocities, are not always and all present. Some bands may be missing, others are more or less dense. The shape and structure of a summed dust cloud vary in a relatively short time, and depend on the trapping date and the size of its particular dust clouds.

The imaging polarimetric patterns in the 2nd part of this paper (Slíz-Balogh, Barta, Horváth 2018) confirm the structure of the KDC shown in Figure 5B. This remarkable similarity may mean that there was no significant particle trapping 6 or 7 or 19 or 20 days before the date of the polarimetric measurement.

The two kinds of summed dust clouds seen in Figures 5B and 5C show an interesting match with the two types of *Gegenschein* described by (Moulton 1900): (i) a large and round (Figure 5B), or (ii) a very much elongated (Figure 5C), varying in a few days time scale, similarly to our simulations.

Our simulations showed that the dust particles trapped earlier than 20-25 days do not contribute to the dust cloud's structure, because after that time the dust is smoothly distributed. This also means that if we see bands, they are the results of trappings not earlier than 20-25 days.

We assume that our simulated particle conglomeratum (summed dust cloud) around the L5 point (Figure 5) corresponds to the dust cloud photographed by Kordylewski (1961). Salnikova *et al.* (2018) presented another computer model of the dust cloud formation around L5, and they also concluded that the accumulation of dust particles is indeed possible around L5.

Although our simulations were performed with a negligible (relative to the primaries) mass of the test particle, the results are the same for particles as massive as $10^6$ kg (Slíz *et al.* 2015), if we disregard the gravitational interaction between the test particles. This means that even rock-sized objects can circulate along with the Lagrange point L5 for a long time.

The observation of the KDC with imaging polarimetry is much reliable than that with photometry. Thus, it is imaginable that the KDC did not reveal itself in the infrared patterns measured by IRAS (https://lambda.gsfc.nasa.gov/product/iras/docs/exp.sup/toc.html) and COBE (https://science.nasa.gov/missions/cobe), especially if astronomers did not search it directly. Furthermore, since longer wavelengthes are scattered less than shorter ones and the KDC can be detected by dust-scattered light, the photometric detection of the KDC is more difficult in the infrared than in the visible spectral range. Finally, the lack of photometric detection of the KDC by earlier astronomical missions (e.g. IRAS, COBE) does not exclude at all the existence of this dust cloud detected by us with imaging polarimetry (Slíz-Balogh, Barta, Horváth 2018). Note that the major aim of all earlier photometric missions was quite different than the detection of the KDC. If during the evaluation of the registered photometric patterns of these missions researchers did not look directly for the KDC, then the chance of its detection was considerably reduced, if not zero.

Similar happened with the detection of the 4th polarizationally neutral point of the Earth atmosphere: The existence of this neutral point was predicted by David Brewster in the 1840s, after his discovery of the 3rd neutral point, named after its first observer, Brewster (1842). However, the 4th neutral point can be observed only from higher altitudes (> 1 km from the Earth surface), which limitation made difficult such an observation. Thus, the first scientifically documented observation of the 4th neutral point happened only in 2002 (Horváth *et al.* 2002). Interestingly, in 2002 the satellite-born imaging polarimeter, called POLDER (Deschamps *et al.* 1994) was already registering the polarization patterns of earthlight for several years. The polarization traces of the 4th neutral point should also exist in the polarimetric data of the POLDER mission. In spite of this, POLDER researchers did not recognize the 4th neutral point, because they did not seek it; they were interested in quite other aspects and meteorological applications of the POLDER-measured polarization data. However, if POLDER researchers have looked for the 4th neutral point, they surely would have found it in their polarization patterns measured from the high altitude of the POLDER satellite, as Horváth *et al.* (2002) found it in their polarization patterns measured from 3.5 km from a hot air balloon.

**Author Contributions**
Substantial contributions to conception and design: JSB, AB, GH
Performing experiments and data acquisition: JSB, AB
Data analysis and interpretation: JSB, GH
Drafting the article or revising it critically for important intellectual content: JSB, GH

**Acknowledgements:** We are grateful to Miklós Slíz (software engineer, Graphisoft, Budapest, Hungary) for the development of the computer simulation software. We thank Tatiana Salnikova for her valuable comments on an eralier version of our manuscript.

**Competing interests:** We have no competing interests.

**Funding:** No funding is declared.